\documentclass[runningheads]{llncs}
\newcommand{\IGNORE}[1]{}

\usepackage{graphicx}
\usepackage{amsfonts}
\usepackage{amsmath}
\usepackage{algorithm}
\usepackage{algpseudocode}
\usepackage{stix}
\usepackage{xcolor}
\usepackage{comment}
\usepackage{float}
\begin{document}
\title{Maximizing Energy Battery Efficiency in Swarm Robotics
}

\author{Anthony Chen
\and John Harwell
\and Maria Gini
}
\institute{Department of Computer Science and Engineering, University of Minnesota\\
\email{\{chen4714,harwe006,gini\}@umn.edu}}

\maketitle

\begin{abstract}

  Miniaturization and cost, two of the main attractive factors of swarm robotics,
  have motivated its use as a solution in object collecting tasks, search \& rescue
  missions, and other applications. However, in the current literature only a few papers consider energy allocation efficiency within a swarm. Generally,
  robots recharge to their maximum level every time unconditionally, and do not
  incorporate estimates of the energy needed for their next task. In this paper we present an energy
  efficiency maximization method that minimizes the overall energy cost within
  a swarm while simultaneously maximizing swarm performance on an object gathering
  task. The method utilizes dynamic thresholds for upper and lower battery
  limits. This method has also shown to improve the efficiency of existing energy management methods.

\end{abstract}


\section{Introduction}

Swarm intelligence has been developed through the observation of cooperation in natural
swarms, such as fishes, birds, and bacteria~\cite{marco}. In the context of a
foraging task, ant colonies have been studied as they can execute simple, efficient,
localized foraging algorithms that result in an intelligent division of labor based on
assigning roles according to past performance~\cite{khaldi}. Ants can also communicate
with one another through pheromones, through which they achieve an advantageous collective foraging
 behavior. Similarly, in swarm robotics, virtual methods to imitate
pheromones have been simulated to communicate between robots~\cite{payton}.
In general, localized communication is necessary to ensure system scalability to
thousands of robots \cite{khaldi}.

A common application in swarm robotics is foraging, in which robots search widely for
resources to bring back to a central location (the ``nest'').  When comparing
foraging methods to determine which one is most suited to a particular application,
some of the desired features are simplicity, scalability, decentralization, sensing,
and parallelism~\cite{Liekna}. Systems with these properties are much better equipped
to handle applications with multiple objectives in dynamic environments, such as
post-disaster relief and geological surveying \cite{TAN}.

In most prior work, the metric for efficiency is the time spent foraging~\cite{EnergyTimeEI}. Time spent foraging does
not fully encapsulate the cost 
it takes to perform the task, because it
ignores actual energy used or stored, as well as the time taken to re-charge a given
robot.  Although it is important for swarms to maintain an optimized speed and
efficiency~\cite{TAN}, time is not the only metric for energy usage.
Another measure is the remaining battery residing in each robot. It is
important to maintain an efficient use of the energy in the robots as well as the energy
being depleted while foraging. If robots were charged to 100\% without consideration
to how much energy is actually needed, the swarm would be wasteful. It would be wasteful to
charge robots to their maximum capacity if it was already known that there were resources near
the nest. That robot could use its remaining energy to perform those simple
tasks. Other robots that are struggling to find resources in other areas would be
the ones that have to pay attention to getting charged and using more of their maximum capacity. 

Very few papers consider the energy in the battery when measuring energy efficiency, so when they try to move the algorithms to the real world, their algorithms might be impractical.
Previous work~\cite{WLiu,Stirling,Labella} only focused on measuring energy during
foraging, and does not account for the unused energy in each robot's battery. That energy is wasted for the task 
and therefore the swarm is less efficient than an energy-aware swarm.  The most
commonly used metric for cost is search and retrieval time; 
an additional useful metric
would include the measurement of the energy remaining in each robot's battery.

Regardless of performance on a foraging task, systems need to be energy aware, as the
``income'' from foraging (i.e., how many objects are gathered in a given amount of
time) needs to outweigh the cost of obtaining it~\cite{campo_dorigo}. In this work, cost is
measured in terms of swarm energy consumption. 

As a motivating example, consider a
swarm of robots in which each robot has a high accuracy image recognition camera,
making the swarm as a whole capable of highly precise localization and mapping. Real
time image recognition is highly computationally expensive while the camera hardware
attached to each robot is physically expensive. Such swarm would necessarily
consume a high amount of energy per-capita, potentially making its usage
intractable~\cite{haverinen}. Furthermore, the production of hundreds of thousands of
such robots would likely be sub-optimal in terms of the increased ``income'' received
during foraging vs the additional energy expenditure.

Some of the most popular applications of swarm-robotic foraging are in rescue
missions and agricultural foraging~\cite{Liekna}. Such tasks require an
efficient use of energy as unnecessary foraging may deplete energy that could instead be used
in other tasks. For example, energy savings achieved during foraging could be used
later by the swarms to further process resources if there was a finite amount of
energy available for swarms to use for recharging. In addition, in cases of
non-uniform object distributions, robots in areas with a lot of easy to access
resources would not need to be charged as often. Non-uniform robot charging would
enable robots to be charged with the necessary energy for their next foraging
task. Collectively the swarm would then be more flexible and robust as it can invest
energy where it is most needed.

In this paper, we introduce a method in which each robot stores energy thresholds and
capacity variables to indicate how much energy should be allocated and used during
foraging. In our proposed approach, these variables adapt based on each robot's
environmental encounters (i.e., obstacle collisions, resource pickup, etc) every time
the robot returns to the nest after foraging. Using this method, the swarm obtains a
collective foraging strategy that maximizes both overall energy efficiency and
successful income of resources simultaneously. To validate the effectiveness of our
method, we conducted experiments that investigated energy and performance patterns over
time as the amount of energy consumed by each robot changes depending on the
environment. 
Our results show we can reach our goals and be better than prior work
because of the flexibility of our method, which can be used to improve the efficiency of existing methods, including other rivaling energy efficiency methods.

\section{Related Work} \label{relatedworks}

Liu focused on energy efficiency by changing the foraging time for each robot
\cite{WLiu}. The time spent foraging for resources depended on various cues such as
personal successful food retrievals, collisions with teammates for food, and success
among other robots in food retrieval. So the method was able to find an adaptable and optimal time for foraging, but did not optimize the amount of energy allocated for it.

Stirling took a novel approach at finding energy efficient searching algorithms for
flying swarms in indoor environments~\cite{Stirling}. The strategy involves having
robots create a network of beacons that communicate with each other to direct where
other robots in the swarm should go. When an exploring robot arrives at an unexplored
location, it becomes a beacon to help sense for the other exploring robots. They
found that launching the robots incrementally rather than all at once decreased total
energy consumption as well as collision rate but increased search time. Their method
provides a trade-off between energy consumption and search time. Once again the metric is energy consumption and not energy allocation so it ignores the unused charge in each robot after the task was done.

Labella took inspiration from ants. He created an adaptation method
that controls the number of robots foraging in the environment~\cite{Labella}. Each
robot has a probability variable that increases and decreases based on the
number of successes and failures the robot had when foraging. The probability variable
dictates the probability that the specific robot would leave the nest and start
foraging. This allows for robots who consistently find and retrieve prey to keep
foraging, while also eventually reaching an equilibrium or state in which the
necessary number of robots are active. Thus, it efficiently uses energy when
needed, similarly to~\cite{WLiu,Stirling} in that they measure the time it takes to finish foraging but not how much energy was already allocated. 

\section{Proposed Method}

\subsection{Problem Statement}

We assume \textit{K} foraging robots initialize at a nest. The environment outside
 the nest has a finite \textit{R} number of resources.  Each of the \textit{K}
robots needs to forage for one of these \textit{R} resources and bring it back to the
nest without using more energy than needed. The robot can charge at the nest as the nest has an infinite source of energy. If the robot's battery level reaches 0,
it will be permanently lost. The total time, $t$, it takes for a robot to complete a
single round of foraging from leaving the nest to returning to the nest can be broken
down into time while \textbf{searching}, $t_s$, and time while \textbf{retreating},
$t_r$. The energy level $E$ decreases at different rates depending on whether the
robot is searching or retreating. So the overall change in energy level for a single robot k during $t$ is modeled as:

\begin{equation}
   \Delta E_k = -\underbrace{\alpha_{s} \cdot t_s}_{\text{Energy spent Searching}} - \underbrace{p \cdot f(r)}_{\text{Energy spent collecting if successful}} - \underbrace{\alpha_{r} \cdot t_r}_{\text{Energy spent Retreating}}
\end{equation}

\noindent where $\alpha_{s}$ is the energy lost rate for searching and $\alpha_{r}$
is the energy lost rate for retreating. $p$ is the energy cost for finding and
collecting a resource, $r$, while $f(r)$ is a binary function that returns if 1 or 0
upon successful or unsuccessful resource retrieval.

Collectively, the swarm solves the following multi-objective optimization problem:

\begin{equation}
  \max_{f(r)}\min_{E}{\sum_{k\in{K}}\Delta E_k{f_k(r)}}
 \end{equation}

We want to maximize the amount of successful resource retrievals while also minimizing the energy allocation of every robot.

\subsection{Adaptive Energy Parameters}

With $\Delta E_k$ we denote the change in the energy level of a robot $k$ after a
round of foraging. We now introduce how the battery energy is partitioned for use in the
next round.

\begin{equation}
    \textit{Energy allocated} = \textit{Energy to Forage} + \textit{Energy to Retreat}
\end{equation}

\begin{equation}
    E^U(t) = C + E^L(t)
\end{equation}

First, energy is allocated for foraging resources (searching + collecting). Second,
energy is allocated for travelling back to the nest to ensure a safe return.

The energy allocated for foraging resources will be represented by a capacity $C$. The
size of the capacity is described by the difference between an upper threshold,
$E^U(t)$, and lower threshold, $E^L(t)$ value. If the energy level $E$ reaches below
$E^L(t)$, then the energy allocated for retreating back to the nest is
used. Therefore, the amount of energy allocated to return to nest will be determined by what
level $E^L(t)$ is at.

\begin{figure}[h!]
  \centering
  \includegraphics[scale=0.3]{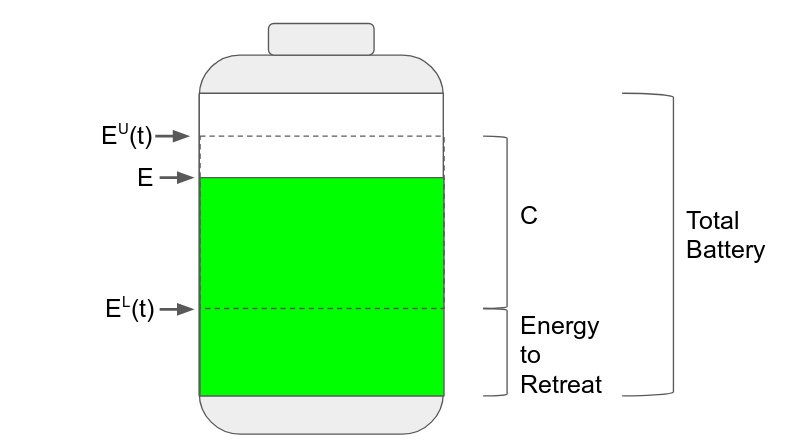}
  \caption{Diagram of battery and battery variables}
  \label{fig:ants}
\end{figure}

The amount of energy for foraging and the amount of energy for retreating needs to be
adaptable to changes in the environment through the changes of values $C$ and
$E^L(t)$, respectively. To ensure energy is allocated properly for the task, both
values will be changed based on successful foraging $f(r)$, $v$ number of robots
encountered, and speed of task completion. 

\begin{equation} \label{eq:4}
     \Delta C = -\underbrace{f(r)\cdot max(0, C - \Delta E)w_{1c}}_{\text{If robot finishes too early}}
     + \underbrace{\neg f(r)w_{2c}}_{\text{If no resource found}}
     + \underbrace{vw_{3c}}_{\text{If encountered other bots}}
\end{equation}

Each of these three components is accompanied by weight variables $w_{1c}$,$w_{2c}$, and $w_{3c}$. When the environment size is held fixed, 
$C$ should increase when (1) there are many robots in the environment and (2)
there are few resources in the environment. It should then decrease when the opposite
is true. A high robot density in an environment means there is more competition among
robots for resources so more energy is needed to look for those resources.
The competition for resources is measured by collision encounters with other robots.
Similarly, few resources also means longer time to forage and so more energy is needed. An
increase in foraging also results in farther distance from the nest which requires
$E^L(t)$ to change similarly:

\begin{equation} \label{eq:5}
    \Delta E^L(t) = -\underbrace{max(0, C - \Delta E)w_1}_{\text{If robot finishes too early}}
     + \underbrace{\neg f(r)w_2}_{\text{If no resource found}}
     + \underbrace{vw_3}_{\text{If encountered other bots}}
\end{equation}
Each of these three components is accompanied by weight variables $w_{1}$,$w_{2}$, and $w_{3}$.
The new upper energy threshold $E^U(t)$ can then be determined by $C$ and $E^L(t)$.

\begin{equation}
    E^U(t) = E^L(t) + C
\end{equation}

A clear emergent property of our proposed algorithm is per-robot energy allocation
such that upon return to the nest a robot $k$ has very little energy left. This is desirable because it indicates that very little wasted energy is used and we efficiently allocate only what is needed for the task. So even if the energy source gets cut off or taken away, each robot should be able to finish their foraging task with little wasted energy.

\subsection{Finite State Machine Implementation}
Our proposed mathematical model is implemented with the help of a per-robot state
machine, described below. A robot $k$ can be in one of 4 states:: \textbf{Charging},
\textbf{Searching}, \textbf{Collecting}, and \textbf{Retreating}.

\begin{equation}
    state = \begin{cases}
                \mbox{searching,} & \mbox{if} E = E^U(t) \parallel f(r) = 0 \\
                \mbox{collecting,} & \mbox{if robot encounters resource} \\
                \mbox{retreating,} & \mbox{if} E <= E^L(t) \parallel f(r) = 1 \\
                \mbox{charging,} & \mbox{if k is in the nest}
            \end{cases}
\end{equation}
\begin{center}
    \text{\textbf{Eq. 7.} State Transitions based on Energy Levels/Thresholds}
\end{center}

The robot starts in \textbf{Searching} from the nest until it finds a resource. Once
a resource is found, the robot tries \textbf{Collecting} it. If it succeeds, it
begins \textbf{Retreating} to nest with it. If not, it returns to \textbf{Searching}
for another resource. If while \textbf{Searching}, the robot's energy level is below
the lower energy capacity threshold $E^L(t)$, then it starts \textbf{Retreating}
without a resource. Once the robot has retreated back to the nest, it starts
\textbf{Charging} until the energy level is at the desired upper energy capacity
threshold $E^U(t)$. Once fully charged, the robot starts \textbf{Searching} and a new
round of foraging begins. 
A nest delay timer is added to the robot while it is charging to simulate the time delay it takes to get a robot charged and ready.

\begin{figure}[h!]
  \centering
  \includegraphics[scale=0.27]{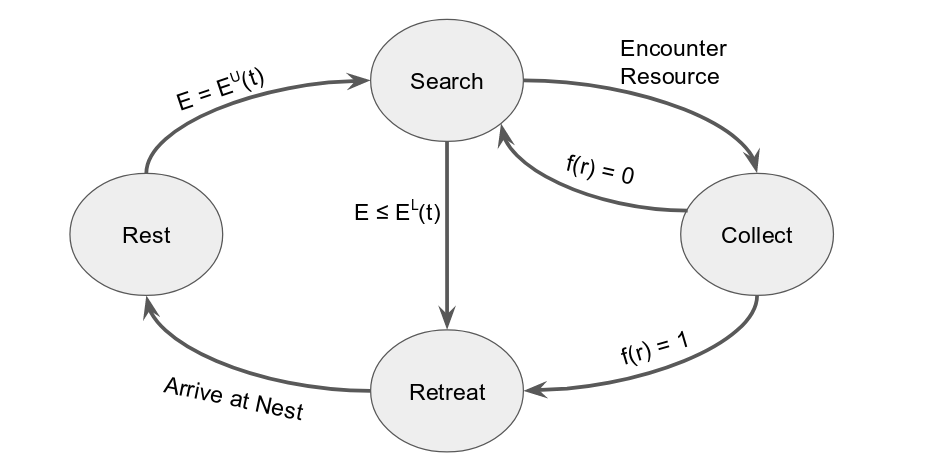}
  \caption{Diagram of finite state machine of behavior for a battery efficient swarm}
  \label{fig:ants}
\end{figure}

How each robot runs each round of foraging and cycle of states for a robot is defined
in the Adapting Battery Algorithm (see Algorithm  \ref{alg:adapt}). The algorithm also handles the adaptive behavior
of the energy capacity and thresholds while the robot is in the nest.

\begin{algorithm}
 \caption{Adapting Battery Algorithm \label{alg:adapt}}
 \begin{algorithmic}[1]
 \State $ \textrm{robot k is initialized with thresholds} E^L(t) \textrm{and} E^U(t).$
 \State $E \gets E^U(t)$
 \State $f(r) \gets 0$
 \State $v \gets 0$
 \State nest delay $\tau_n \gets 20$
 \While {$ R > 0 \parallel \textrm{all K robots inactive or dead} $}
 \If {$E == E^U(t)$}
 \State $state \gets searching$
 \ElsIf {$state == charging$}
    \If {$\tau_n > 0$}
        \State $\tau_n \gets \tau_n - 1$
    \Else
        \State $E^L(t) \gets E^L(t) -max(0, C - \Delta E)w_1 + \neg f(r)w_2 + vw_3$
        \State $C \gets C -f(r)\cdot max(0, C - \Delta E)w_{1C} + \neg f(r)w_{2C} + vw_{3C}$
        \State $E^U(t) \gets E^L(t) + C$
        \State $E \gets E^U(t)$
        \State $f(r) \gets 0$
        \State $v \gets 0$
    \EndIf
 \ElsIf {$state == retreating$}
    \State $E \gets E - \alpha_{r}$
 \ElsIf {$state == searching$}
    \State $E \gets E - \alpha_{s}$
 \ElsIf {$E < E^L(t) \textrm{ and } state = searching$}
    \State $state \gets retreating$
 \ElsIf {Encounter resource r}
    \State $state \gets retreating$
    \State $f(r) \gets 1$
    \State $E = E - p$
 \ElsIf {Encounter robot}
    \State $v \gets v + 1$
 \ElsIf {Encounter Nest}
    \State $state = charging$
    \State $\tau_n \gets 20$
 \EndIf
 \EndWhile
 \end{algorithmic}
\end{algorithm}

\subsection{Adaptive Battery Critical Point}
At some point, the adaptive thresholds of the battery may 
increase to cover the
entirety of the battery. This is because as time passes, most of the resources will
have been foraged while the number of robots remains the same. To find the remaining
resources, the robots will have to use their maximum battery capacity to forage. At
this point, the robot cannot increase its thresholds any more which means it is
near the end of foraging most of the resources. The only threshold that can be
modified is the energy threshold for when to retreat. This threshold will update
similarly to Equation \ref{eq:4} and Equation \ref{eq:5}. We call this period Energy
Efficiency Endgame (EEE).  We investigate three different options for how each robot
will handle being in this state.

\subsubsection{Well-informed.}

In this method, we assume that each robot is well-informed. This means that each robot no
longer needs to change the thresholds of the battery for when it needs to
retreat. Instead, each robot will simply continue but with an increasing nest delay
timer each time the robot re-enters the nest. The nest delay timer increases by $\tau$
seconds after each round of foraging. 

\subsubsection{Ill-informed.}
In the ill-informed method, the robot believes that it has autonomously decided on a
wrong threshold for when to retreat to the nest. So along with increasing the nest
delay timer, the robot also continues to adapt the lower energy threshold using equation \ref{eq:5}.

\subsubsection{Null-informed.}
Null-informed is an option that makes each robot stop foraging and remain still
in the nest. Once the robot has reached its battery full capacity, it stops and relies on
the still adapting robots to find the remaining resources.

\begin{table}[h]
\begin{center}
\caption{Table of Energy
Efficiency Endgame (EEE) options
\label{table_example}}
\begin{tabular}{|c|c|c|c|}
\hline 
EEE Option & Increase nest delay & Adaptive & Stops When \\
\hline \hline
Well-informed & Yes & No & No robot is foraging\\
\hline
Ill-informed & Yes & Yes & No robot is foraging\\
\hline
Null-informed & No & No & Each robot reaches EEE\\
\hline
\end{tabular}
\end{center}
\end{table}

\subsection{Experimental Framework}\label{sec:experiment}

The experiments in this paper were implemented and executed using the ARGoS
simulator, which allows real-time simulation of large swarms of robots \cite{argos}. The
simulation is tested in a three-dimensional space with s-bot models.

\begin{figure}[b]
  \centering
  \includegraphics[scale=0.15]{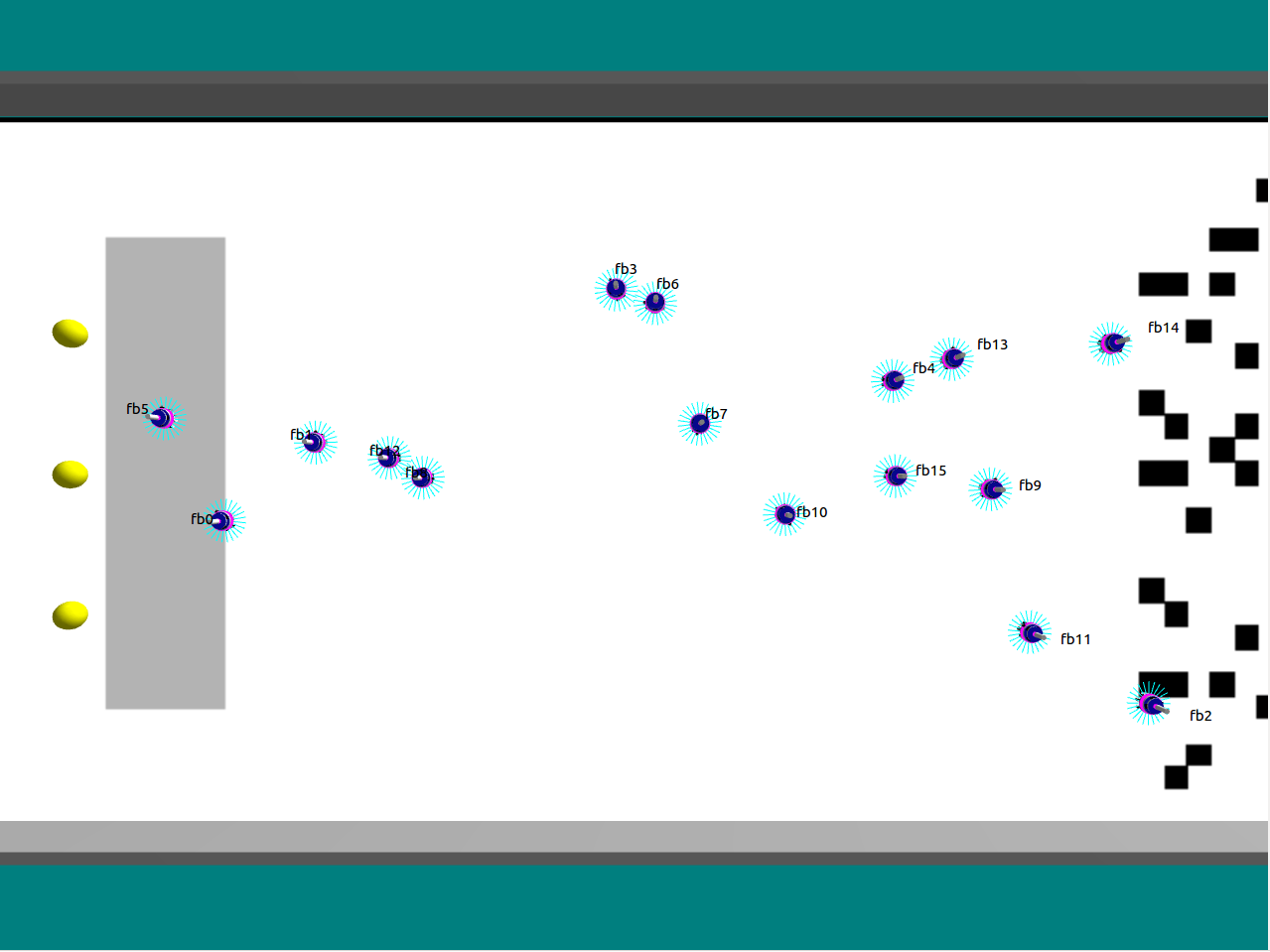}
  \caption{Visualization of the simulation with 16 robots.}
  \label{fig:graph1}
\end{figure}

The following assumptions were made for all experiments:
\begin{itemize}
    \item The robots are homogeneous.
    \item The robots cannot communicate.
    \item The number of resources does not affect a robot's performance.
    \item The environment is flat and open with no unknown objects/obstacles.
    \item The nest has unlimited storage capacity.
    \item Robots can be charged to a specific amount.
    \item The resources in the environment respawn until at least 100 are collected. At that point, 25 resources will remain.
    \item The energy expenditure rates for searching $\alpha_{s}$ and retreating $\alpha_{r}$ are constant.
    \item The energy expenditure $\Delta E$ of a robot can be accurately measured and recorded
    \item All resources are identical.
\end{itemize}

The goal is to find a better method for energy efficiency. Let $r$ be the number of
collected resources and $t$ be the total foraging time. The
\textit{efficiency} of a swarm is often defined as:

\begin{equation}
    \eta = \frac{r}{t}
\end{equation}

\noindent However, in this work, we also consider the remaining energy in each
battery upon task completion as part of our efficiency definition (i.e., robots try to
ensure that the foraging ends with no leftover energy). Thus, our \textit{efficiency} is represented as:

\begin{equation} \label{eq:eta}
    \eta' = \frac{r}{\sum_{k}^{K} E^k_d + E^k_b}
\end{equation}

where $K$ denotes all robots, $k$
denotes a single robot instance, $E^{k}_d$ is energy depleted and $E^k_b$ is energy
remaining in battery of robot $k$. 

Table \ref{valtable} shows the values for the adapting equation that we used in our simulations.
These values were empirically chosen to balance exploration and exploitation. For example, the values $w_{1}$ and $w_{1C}$ were chosen especially small as they represent the coefficient for number of robot collisions. Because a robot could experience many collisions in just one foraging round, we wanted this weight to be small to reduce noise and prevent random collisions to overshadow the other parts of the model. 
\begin{table}[h]
\caption{Values used in the experiments
\label{valtable}}
\begin{center}
\begin{tabular}{|c|c|c|c|c|c|c|c|c|c|}
\hline 
 Init $E^L(t)$ & Init $C$ & $w_1$ &$w_2$ & $w_3$ & $w_{1C}$ & $w_{2C}$ & $w_{3C}$ & $\tau$\\ [0.5ex] 
\hline \hline
 0.3 & 0.5 & 0.3 & 0.1 & 0.005 & 0.2 & 0.1 & 0.005 & 10 \\ 
\hline
\end{tabular}
\end{center}
\end{table}

\section{Results}

For each method, we experimented with swarm sizes of $s \in \{2^i : i=1\dots8\}$ in order to observe efficiency across scales. Each experimental run was terminated when the swarm successfully collected all the resources or the swarm fulfilled the Energy Efficiency Endgame stopping criteria. Performance is measured by $\eta$ as described in Eq. (\ref{eq:eta}). For each experiment, 20 simulations were averaged.
\begin{figure}[ht]
  \centering
  \includegraphics[scale=0.5]{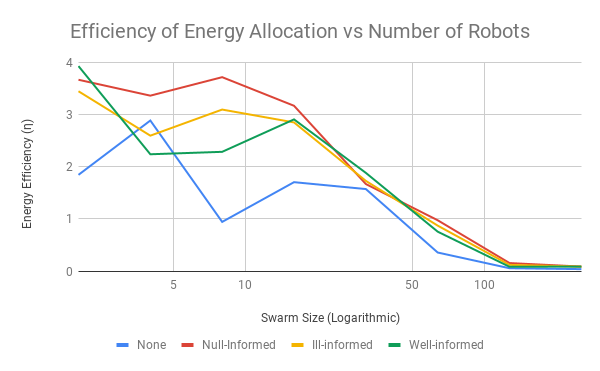}
  \caption{Graph of the efficiency of the different battery allocation methods}
  \label{fig:graph1}
\end{figure}
In Fig.\ref{fig:graph1}, we can see the battery energy allocation method plotted with three different lines each representing a different EEE method. A fourth line-plot was added that forages without any energy efficiency strategy to represent the baseline to show how our method improves efficiency. The results indicate that each energy battery allocation method improved overall swarm efficiency with the Null-informed EEE method raising overall efficiency consistently the most with a variety of swarm sizes. The Null-informed version most likely performed the best because it prevented robots from re-entering the environment when only a few robots were needed to finish collecting the resources. In some cases we can see that having battery conscious robots in the swarm doubles the energy efficiency use. However, it appears that it is not always consistently the best. Specially as we increase the swarm size, the difference in efficiency between a conscious and non-conscious battery energy swarm decreases. For higher swarm densities, this method is not that advantageous, most likely due to competition among robots to retrieve resources. When competition is high, the entire battery capacity is needed so trying to partition the battery usage is futile. But it is important to note that although it appears that difference between energy efficiency is low, it is still between twice to three times more efficient than without it. So the small difference in energy efficiency improvement is more due to the arena ratio than the efficiency method itself.
\begin{figure}[h]
  \centering
  \includegraphics[scale=0.5]{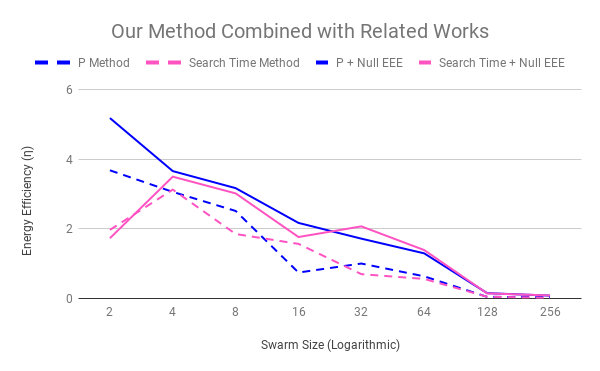}
  \caption{Efficiency of different methods with or without battery allocation methods. Dotted lines (red and blue) are related energy efficiency works that do not use our battery conscious allocation strategy. The solid lines (green and yellow) are those related works methods in conjunction with our method.}
  \label{fig:graph2}
\end{figure}
In a second set of experiments, we explored how the efficiency improves when applying our conscious battery energy allocation method compared to other existing energy efficiency methods. Previously discussed in Section \ref{relatedworks}, Labella's group used an adapting probability value to control when a robot should leave the nest~\cite{Labella}. We implemented the algorithm with the same experimental values $P_{min}$ = 0.0015, $P_{max}$ = 0.05, $P_{init}$ = 0.033, and $\triangle$ = 0.005.

Similarly, Liu's group had an adaptable time variable that denoted how long the robot should spend searching \cite{WLiu}. However, since they used pheromones as social cues for their time variable, we will have to modify it since we are not using pheromones. Instead we simply have the robot keep track its own successes and failures rather than those of other robots. The constant values used from their experiment equation remain the same.

In Fig \ref{fig:graph2}, the dotted lines denote related approaches without our conscious battery energy method and the solid lines denote related approaches with the conscious battery method.  Our method almost consistently increases the energy efficiency in every experiment compared to the other energy efficient methods. This is indicated in the graph by the fact that the solid lines (methods with Null EEE) have higher energy efficiency than the dotted lines (methods without Null EEE). It is also interesting to note that if you compare \ref{fig:graph2} with \ref{fig:graph1}, our battery conscious strategy by itself is about as efficient or almost as efficient as the related works methods. And then combining the two improves efficiency significantly. This shows the advantage of using our conscious battery allocation in future swarm foraging methods.

\section{Conclusions and Future Work}
We presented an adaptive, energy efficient, and energy-aware approach to foraging
swarm robotics. Through conscious allocation of energy our method has shown that energy efficiency can be greatly increased when taking into account the energy allocated in the battery. When using battery conscious efficiency methods such as the Null-informed EEE, swarm strategies increase energy efficiency significantly.

One direction for future work is to further explore how combinations of energy efficient
methods perform. It would be interesting to see if combining our method
with other existing efficiency method would increase efficiency or if too much
complexity will produce a diminishing return. Another aim could be to explore local
communication or signaling among robots to see if there is a way to communicate
energy information. With this, robots could adjust energy allocation and usage with
other robots are taken into account instead of just the robot itself. 

Another direction is to explore multiple nest locations each with limited charging capacity. In real life, each nest for swarm robots is not going to have unlimited energy supply. So a future research topic is to explore energy efficiency when supply is limited and perhaps explore robot communication to direct which nest has more energy supply. 

As listed in the Experimental Framework (Section \ref{sec:experiment})
we assume that the energy expenditure rates for searching $\alpha_{s}$ and retreating $\alpha_{r}$ are constant. We also had our simulation be able to accurately calculate and record the total expenditure $\Delta E$. The issue with these two assumptions is that it is difficult to implement this model on robot hardware due to how accurate the real-time energy measurements must be. One direction we would like to explore is trying to run real life swarm robot experiments that can handle this model. One possibility is having the energy levels be measured in time, so the energy remaining is measure in how many seconds left before the robot dies. This would of course require lots of battery testing to see how the energy expenditure rates change in real time. 

In order to facilitate future research and collaboration, the code for this work is
open source. It can be found at~\texttt{https://github.com/swarm-robotics}.


\section*{Acknowledgments}

 We gratefully acknowledge Amazon Robotics, the MnDRIVE RSAM initiative at the University of Minnesota, and the Minnesota Supercomputing Institute (MSI) for their support of this work.



\bibliography{references}

\end{document}